\documentclass[graybox]{svmult}

\usepackage{ulem}			
\usepackage{mathptmx}       
\usepackage{helvet}         
\usepackage{courier}        
\usepackage{type1cm}        
\usepackage{makeidx}         
\usepackage{graphicx}        
\graphicspath{{Figures/}}
\usepackage{multirow}
\usepackage{multicol}        
\usepackage[bottom]{footmisc}
\usepackage{graphicx}
\usepackage{algorithmicx}
\usepackage{algorithm}
\usepackage{algpseudocode}
\usepackage{amsmath}
\usepackage{cite}
\usepackage{hyperref}
\usepackage{placeins}
\usepackage[font=small]{caption}
\usepackage[labelformat = empty,position=top]{subcaption}
\usepackage[export]{adjustbox}

\usepackage{titlesec}
\titlespacing*{\section}{0pt}{1.1\baselineskip}{\baselineskip}

\usepackage{overpic}

\newcommand{\piclabB}[2]{
\begin{overpic}[width=0.36\linewidth]{#1}
 \put (3,98) {\large \textsf{#2}}
\end{overpic}
}
\newcommand{\piclabC}[2]{
\begin{overpic}[width=\linewidth]{#1}
 \put (3,98) {\large \textsf{#2}}
\end{overpic}
}
\newcommand{\piclabD}[2]{
\begin{overpic}[width=0.4\linewidth]{#1}
 \put (3,98) {\large \textsf{#2}}
\end{overpic}
}
\newcommand{\piclabE}[2]{
\begin{overpic}[width=0.55\linewidth]{#1}
 \put (3,98) {\large \textsf{#2}}
\end{overpic}
}

\newcommand{\piclabG}[2]{
\begin{overpic}[width=0.23\linewidth]{#1}
 \put (3,98) {\large \textsf{#2}}
\end{overpic}
}
\newcommand{\piclabH}[2]{
\begin{overpic}[width=0.49\linewidth]{#1}
 \put (3,98) {\large \textsf{#2}}
\end{overpic}
}

\makeindex             

\begin{document}

\title*{Inferring short-term volatility indicators from the Bitcoin blockchain}
\author{Nino Antulov-Fantulin*, Dijana Tolic*, Matija Piskorec, Zhang Ce, Irena Vodenska\\ *shared first authorship}

\authorrunning{Antulov-Fantulin, Tolic, Piskorec, Ce, Vodenska}

\institute{
Nino Antulov-Fantulin* \at Computational Social Science, ETH Zurich, Switzerland, \email{anino@ethz.ch}
\and Dijana Tolic* and Matija Piskorec \at Laboratory for Machine Learning and Knowledge Representations, Institute Ruder Bo{\v s}kovi\'c, Croatia
\and Zhang Ce \at Systems Group, ETH, Zurich, Switzerland
\and Irena Vodenska \at Metropolitan College, Boston University, USA
}
%
%

\titlerunning{Inferring short-term volatility indicators from Bitcoin blockchain}
\maketitle
\abstract{In this paper, we study the possibility of inferring early warning indicators (EWIs) for periods of extreme bitcoin price volatility using features obtained from Bitcoin daily transaction graphs.
We infer the low-dimensional representations of transaction graphs in the time period from 2012 to 2017 using Bitcoin blockchain, and demonstrate how these representations can be used to predict extreme price volatility events.
Our EWI, which is obtained with a non-negative decomposition, contains more predictive information than those obtained with singular value decomposition or scalar value of the total Bitcoin transaction volume.   }

\section{Introduction}

Blockchain as a new technology has a potential to change the traditional way of communication, contracting, and financial management. 
The first and still most popular use of blockchain technology is its use as a digital currency, or cryptocurrency, as a part of the the Bitcoin protocol \cite{Nakamoto2008}.
There the payments are processed by a peer-to-peer Bitcoin network where users announce new transactions and which are verified by network nodes and recorded in a blockchain - a public distributed ledger.
Beyond its usage in cryptocurrencies, blockchain technology's essential importance is to offer a new way to record and store confidential information. 
It has a potential to enable services that we do not even consider today, for example to offer support for liberalist's way of decision-making, aid in development of a fair-value, decentralized marketplaces and help increase financial inclusion in developing countries. 
Blockchain could be one of the future solutions \cite{Dirk2016,Future4.0} to secure liberalism and preserve the integrity of policy-making decisions as it promises faster and cost-efficient methods for election voting, as well as protection against manipulations and cyber-attacks .
Blockchain can also be used as a building block for new decentralized marketplaces that offer avoidance of overpricing and manipulation because they provide a place for negotiating contracts that are based on realistic supply and demand in the market.
Blockchain could provide access to financial sector services such as loans, insurance, savings, signing contracts and sending and receiving payments to low-income or socially excluded people in the developing countries \cite{Watanagase15}. This can be achieved solely through mobile internet access, which is more cost-effective than developing traditional financial infrastructure.


Blockchain offers a unique view into today's economic and financial systems that are global and interdependent. Researchers have used network approach to study such complex an dynamic systems to reveal important system characteristics or shed light on inherent network vulnerabilities \cite{acemoglu2013network, huang2013cascading, sakamoto2017systemic, glasserman2015likely, battiston2016complexity, piskorec2014cohesiveness, huang2011identifying, vodenska2016interdependencies}.
This approach is especially appropriate for studying in the Bitcoin network, which connects its users on a global scale and allows them to exchange non-physical, non-regulated, and decentralized financial assets without any economical equivalent or guarantee by a central bank or a sovereign \cite{Yermack2015,AssetCurrency}.
Although it is a relatively new system, different aspects of the Bitcoin have already been extensively analyzed, including price formation \cite{Garcia2014,Garcia2015, amjad2017trading, DidierSpencer}, price fluctuations \cite{Tian, Kim2016, KondorBTC}, systems dynamics \cite{BTCstructure, KondorBTC, ElBahrawy2017}, economic value \cite{Bolt2016, Hayes2015, Kristoufek2015}, limit order book dynamics \cite{BouchaudBTC, Tian}, privacy and security \cite{Shamir, Mser2017}, blockchain protocol and mining process \cite{Garay2015, Eyal2018} and many others. 


In this paper, we are interested in the following question:
Is it possible to infer early warning indicators (EWIs) that are able to predict short-term extreme volatility events on a timescale of 1-10 days, from daily Bitcoin transaction graphs. 
The transaction graphs extracted from blockchain data contain information about the money flow among different Bitcoin addresses without any pricing data. A market price is usually formed as a combination of different complex economic and financial effects \cite{btcEconomy,Bolt2016, btcFormation1, DidierSpencer}. 
According to a recent study \cite{Bolt2016}, the values of virtual currencies are affected by the demand for such currencies to purchase real goods and services, in addition to the speculative buying and selling dynamics on the exchanges. All Bitcoin transactions are written to the blockchain, in form of temporal transaction graphs, where nodes represent different Bitcoin addresses and edges represent the money flow based on transactions (purchases and sales of goods and services). A study by Kondor et. al. \cite{KondorBTC} demonstrated that there exists a certain correlation between the Bitcoin network structure and the market effects i.e. Bitcoin price change, up to early 2014. However, the authors of \cite{KondorBTC} did not test the predictive power on hold-out data.   
This has motivated us to analyze patterns in the transaction graphs from the Bitcoin blockchain using unsupervised and supervised machine learning. Our methodology consists of two main steps: (i) constructing low-dimensional representations of the transaction graphs and (ii) learning how to combine low-dimensional representations in order to be able to predict short-term extreme volatility events. In Section 2, we describe the blockchain data and methods that we use. Section 3 and 4 provide the evaluation of our results and discussion. 

\section{Data and methods}
Blockchain consists out of a list of transactions, each with a certain number of inputs and outputs.
Each input consists of the hash of the current transaction, hash of the previous transaction, the public key of the current input, timestamp, and other data. Similarly, each output has the hash of the current transaction, the public key of the output address, amount of bitcoins, time stamp and other data. 
The user transaction network can be extracted from the blockchain by exploiting the fact that initiating a transaction with multiple inputs requires signing it with the private keys of all input addresses.
This implies that all of these addresses are controlled by the same entity \cite{KondorBTC,Shamir,Nakamoto2008} that we simply call a user.
Similar as in literature \cite{KondorBTC,Shamir}, we process hundreds of gigabytes of Bitcoin blockchain data by merging all addresses that belong to the same user. 
After processing we get the temporal weighted directed transaction networks, where nodes represent users after the merging process. Each link $(i,j,w,t)$ represents a transaction event from source user $i$ to destination user $j$ at time $t$ with $w$ amount of bitcoins. 
We filter only the long-term users that were active before the January 1st 2017.
Users are considered long-term users if they were involved in at least 100 individual transactions and at least 600 days passed between their first and last appearance in the dataset.
This filtering gives us over 106 millions of transactions between 114 768 long-term users which corresponds to over 90\% of all blockchain volume. 
The time evolution of the Bitcoin transaction network is encoded in the matrix $\textbf{X} \in R^{M \times T}$, where $T$ denotes the number of temporal snapshots of a network, that is described with $M$ values. Column $\textbf{x}_{t} \in R^M$ represents encoded temporal snapshot at day $t$. 
In the case of \textbf{edge encoding}, the $i$-th position of vector $\textbf{x}_{t}$ encodes the number of bitcoins that were exchanged through $i$-th edge in day $t$. 
In the case of \textbf{node encoding}, the $i$-th position of vector $\textbf{x}_{t}$ encodes the number of bitcoins that $i$-th node received from all other nodes in day $t$.  

\begin{figure}[t]
\begin{center}
\piclabE{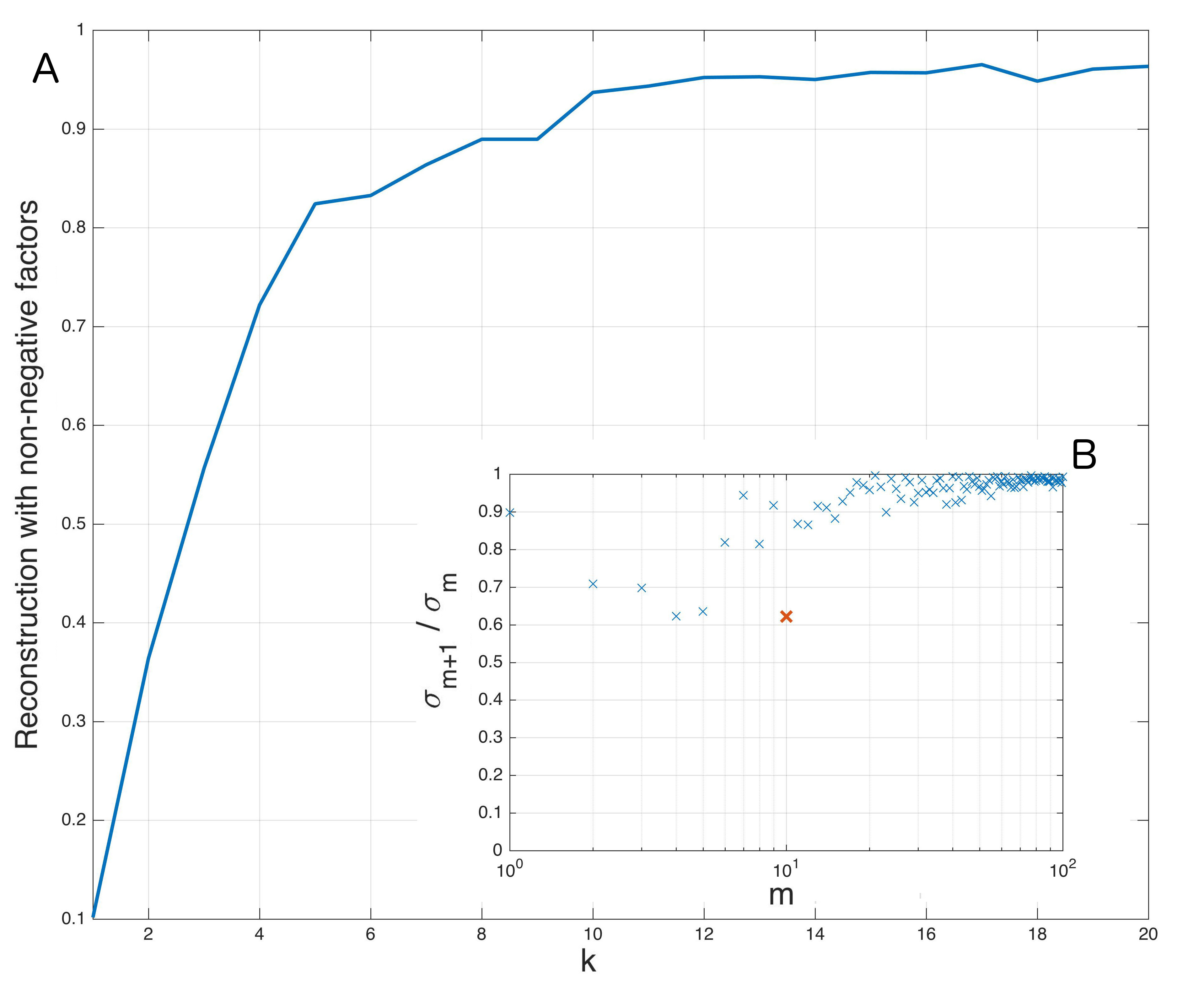}{}
\piclabB{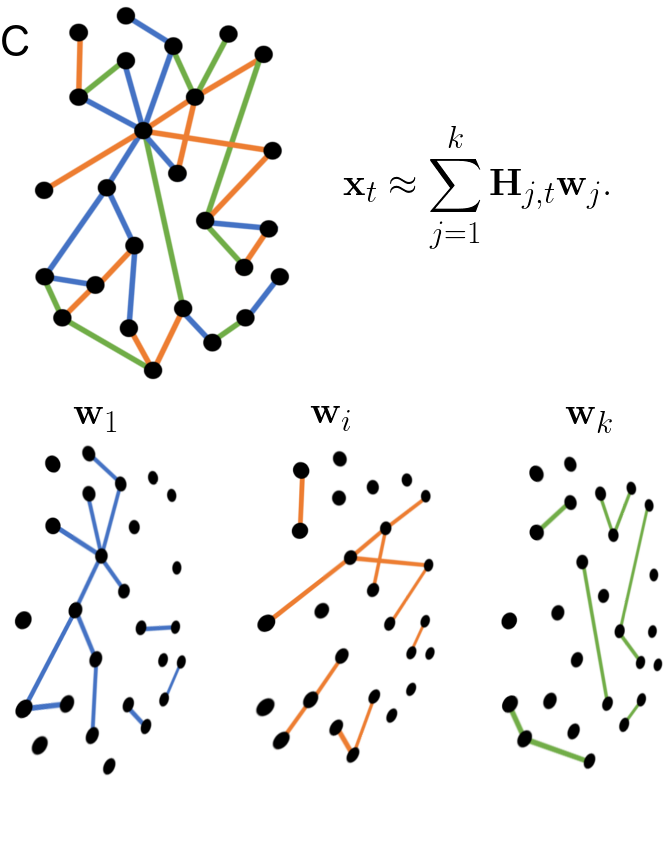}{}
\end{center}
\caption{\textbf{Panel A}: Low-rank reconstruction of the Bitcoin transaction history (edge encoding) with non-negative matrix approximation with rank $k$. The reconstruction is measured as $( ||\textbf{X}||-||\textbf{X}-\textbf{W}_k \textbf{H}_k|| ) / ||\textbf{X}||$, when the reconstruction error $||\textbf{X}||-||\textbf{X}-\textbf{W}_k \textbf{H}_k||$ is zero, the reconstruction is 1. With only 10 base networks we can reconstruct the 95 \% of all the Bitcoin transactions from 2010 to 2013. \textbf{Panel B}: Estimation of the rank of the matrix of the Bitcoin transaction network snapshots. The rank is estimated by finding the index when the ratio between two consecutive singular values is minimum \cite{MatrixCompl}. The estimated rank 10 is indicated by a red cross symbol. \textbf{Panel C:} Visualization of the decomposition of the graph $x_t$ as a linear non-negative combination of base networks $\textbf{w}_j$. Each base network $\textbf{w}_j$ has contribution of $\textbf{H}_{j,t}$ depending on the time $t$.
}
\label{fig:reconstruction}
\end{figure}

\textbf{2.1. Low-dimensional representations of transaction graphs}\\														
We use techniques from unsupervised learning to create low dimensional representations $\textbf{x}_{t} \rightarrow \textbf{h}_{t}$ of the Bitcoin transaction graphs. We employ the non-negative matrix factorization (NMF), which is particularly suited for our problem because it produces non-negative factors which have a clear interpretation - the factors correspond to the (potentially overlapping) subnetworks of the original transaction graph.
This is in contrast to some other matrix factorization methods, for example Singular Value Decomposition (SVD), which can produce factors with negative weights~\cite{KondorBTC}.
The Bitcoin evolution matrix $\textbf{X}$ can be factorized into two non-negative matrices $\textbf{X} \approx \textbf{W} \textbf{H}$,  where the  $\textbf{W}= [\textbf{w}_1, ..., \textbf{w}_k] \in R^{Mxk}$ consists out of $k$ basis vectors $\in R^M$, each of which corresponds to the subnetworks of the transaction graph. The matrix $\textbf{H} \in R^{kxT}$ contains $T$ low dimensional representations of transaction graphs.
Note that we use different terminology depending on the context: basis vectors for linear algebra context, factors for matrix factorization context and base networks for network context.
The reconstruction of the transaction network for day $t$ is the non-negative linear combination of non-negative basis networks:
\begin{equation}
\textbf{x}_t \approx \sum_{j=1}^{k} \textbf{H}_{j,t} \textbf{w}_j.
\end{equation}
This means that each transaction network can be decomposed as a superposition of the transaction subnetworks $\textbf{w}_j$ (see Figure \ref{fig:reconstruction}, panel C), where each contributes with weights $\textbf{H}_{j,t}$. 
We formulate the optimization problem, where we seek NMF factors that minimize the reconstruction error (see Figure \ref{fig:reconstruction}, panel A-B).
In order to handle high dimensional noisy data and outliers, we use the robust NMF \cite{RNMF} formulation:
$
\underset{\textbf{H,W} \geq 0}{\textrm{min}} ||\textbf{X}-\textbf{WH}||_{2,1},
$
where $||.||_{2,1}$ denotes the $L_{2,1}$ matrix norm. 
This norm \cite{L1} is robust to outliers and it is defined as:
$
||\textbf{X}||_{2,1} = \sum_i \sqrt{ \sum_j \textbf{X}_{i,j}^2 } = \sum_i ||x_i||_2,
$
where $||.||_2$ denotes the L-2 norm. 
In order to have sparse representations, we also add the $L_{2,1}$ norm on the encoding matrix $\textbf{H}$ to the optimization function:
\begin{equation}
\label{eq:fobustNMF}
\underset{\textbf{H,W} \geq 0}{\textrm{min}} ||\textbf{X}-\textbf{WH}||_{2,1} + \lambda ||\textbf{H}||_{2,1}.
\end{equation}

This optimization problem is non-convex and it is solved by adopting the iterative procedure to alternatively fix one of the matrices $(\textbf{W},\textbf{H})$ and then solve the convex problem with multiplicative update rules \cite{RNMF} (see appendix).  

\textbf{2.2. Early warning indicator (EWI)}\\
We denote the early warning indicator as $\eta(t)$ and model it as a linear function of low dimensional representations $\textbf{H}$ of a transaction graph. 
As the volatility has non-negative domain, we construct early warning indicator as a non-negative linear superposition of non-negative elements (features) in the encoding matrix $\textbf{H}$:
\begin{equation}
\label{eq:NMF-NLR}
\eta(t) = \sum_{j=1}^k \sum_{t=1}^{\delta} c_{j,t} \textbf{H}_{j,t} \textit{ s.t. } c_{j,t} \geq 0,
\end{equation}
where $k$ denotes the dimensionality of low-dimensional representation and $\delta$ the auto-regressive order i.e. number of historical days used for prediction. 
In the rest of the text we refer to this supervised model (\ref{eq:NMF-NLR}) as \emph{Linear Non-negative Auto-Regressed NMF model} (NMF-NLR). Next, we need to infer the coefficients $c_{j,t}$ in such a way to be able to predict future volatility. 

\textbf{2.3. Inference step}\\
First, we describe the \textbf{partitioning} of data to train and hold-out parts, as well as inference settings. 
We partition the dataset $\textbf{X}$ with respect to $n$ temporal points $h_1 < h_2 < ... < h_n$ into disjoint hold-out segments $X=[\textbf{X}_{[h_1,h_2]}, ...., \textbf{X}_{[h_{n-1},h_{n}]}]$ such that each segment is $\Delta$ days long.
Now, for each hold-out segment $\textbf{X}_{[h_i,h_{i+1}]}$ we use the previous $M$ days $\textbf{X}_{[h_i-M,h_{i-1}]}$ for training.
For simplicity, each training segment $\textbf{X}_{[h_i-M,h_{i-1}]}$ is denoted as $\textbf{T}_i$ and its 
corresponding validation segment $\textbf{X}_{[h_i,h_{i+1}]}$ as $\textbf{V}_i$.
In summary, we have two different partitions of the data: (i) disjoint hold-out segments $X=[\textbf{V}_1, \textbf{V}_2, ..., \textbf{V}_n]$ and corresponding overlapping segments used for training $X=[\textbf{T}_1, \textbf{T}_2, ..., \textbf{T}_n]$. Each model is trained on $\textbf{T}_i$ segment and validated on $\textbf{V}_i$ segment, where we use $\Delta=30$ days for hold-out segments and $M=5*30$ days for training. 

In \textbf{training phase}, for each training segment $T_i$, we perform feature extraction with the non-negative matrix factorization $\textbf{T}_i = \textbf{WH}$.
Matrices \textbf{W,H} are found by solving the optimization problem in equation~\ref{eq:fobustNMF}, defined in section 2.1. 
Recall, that columns in matrix \textbf{H} are low-dimensional representations (features) of daily transaction graphs.
Then, coefficients $c_{j,t}$ are found by minimizing the square difference between EWI and volatility for next day i.e. $(\eta(t)-\sigma_{t+1})^2$. We inferred the non-negative coefficients $c_{j,t}$ for regularized non-negative linear regression by using the updates rules for sparse non-negative coding \cite{SNMF}. 
Note that the inferred coefficients $\textbf{c}$ for the training segment $\textbf{T}_i$ are associated with the base matrix $\textbf{W}$. If we change the base matrix $\textbf{W}$, the representation $\textbf{H}$ also changes. Therefore, for each training segment $\textbf{T}_i$ the model parameters are $M_i = (\textbf{W}, \textbf{c})$.

In \textbf{validation phase}, for each hold-out segment $\textbf{V}_i$, we use the corresponding model $M_i = (\textbf{W}, \textbf{c})$ from previous adjacent training segment. First, we need to extract representations $\textbf{H}$ that are associated to the learned model $M_i = (\textbf{W}, \textbf{c})$. Representations are found by the following convex optimization problem:
\begin{equation}
\label{eq:NMF-Honly}
\underset{\textbf{H} \geq 0}{\textrm{min}} ||\textbf{V}_i-\textbf{WH}||_{2,1} + \lambda ||\textbf{H}||_{2,1}.
\end{equation}
Note, that the matrix $\textbf{W}$ is fixed and therefore we only use the update rules for finding matrix $\textbf{H}$ (see Appendix). 
Finally, we use the coefficients $\textbf{c}$ to form predictions on hold-out segment with equation \ref{eq:NMF-NLR}.
Fixing a base matrix $\textbf{W}$ is necessary if we want to use the inferred coefficients.

\section{Results} 
Our final aim is to be able to predict short-term extreme volatility events, not the volatility value itself. 
At day $t$ we want to predict that the extreme event will happen in future segment $[t+1,t+1+h]$ of $h$ days. In a special case, when $h=1$ we have a localized prediction for next day. From machine learning perspective, we want to classify future segment into class ``1'' or ``0'', where class ``1'' means extreme volatility event. More formally, based on the EWI $\eta(t)$, we make prediction $\hat{\beta}(.)$ for segment $[t+1,t+1+h]$ as:
 \begin{equation}
\hat{\beta}([t+1,t+1+h]) =
  \begin{cases}
   1  :& \eta(t) \geq \Theta\\
   0  :& \text{else}
  \end{cases}.
\end{equation}
\textbf{3.1. Extreme event definition} 
The price fluctuations are measured with the Garmann-Klass \cite{Garman-Klass} definition of volatility. That is calculated as $\sigma(t)^2 = \textstyle\frac{1}{2}\displaystyle \left( \log (H_i/L_i) \right)^2  - (2\log 2-1) \left( \log (C_i/O_i) \right)^2$, where $O_i, H_i, L_i, C_i$ stand for open/high/low/close daily price. 
If the level of volatility exceeds some threshold $\alpha$, we will consider it as an extreme volatility event.  
We use the following threshold levels $\alpha=\left\lbrace 0.05, 0.1, 0.15, 0.2 \right\rbrace$, which result in 18\%, 5\%, 2.5\% and 1.6\% of events being labeled as extreme ones in period from 2012 to 2017.
A time segment of length $h$ is considered extreme if it contains at least one extreme volatility event, independent on it's localization. One can think of $h$ as a localization parameter in future horizon. 
The ground truth is denoted as $\beta(.)$ and for segment $[t+1,t+1+h]$ of $h$ days in future is:
 \begin{equation}
\beta([t+1,t+1+h]) =
  \begin{cases}
   0  :& \sigma(t) < \alpha: \forall t \in [t+1,t+1+h]\\
   1  :& \text{else}
  \end{cases}
\end{equation}
Simply, if the daily volatility $\sigma(t)$ in next $h$ is always less than $\alpha$, we mark this segment with label ``0''. 
Although our prediction task is classification, we have used the regression in the inference step, which is not  uncommon practice in machine learning \cite{Suykens1999}. 
Remember that the vector $\textbf{x}_t$ denotes the snapshot of Bitcoin network dynamics at day $t$. 
Due to the scalability issues we have used the node encoding, rather than edge encoding, to describe the snapshot of Bitcoin dynamics. The node encoding, on every $i$-th position in vector $\textbf{x}_{t}(i)$ has the value of the total number of Bitcoins that node $i$ received from other nodes during one day $t$. In the future work, we plan to analyze edge encoding version in more detail.

\begin{figure}[t]
\begin{center}
\piclabC{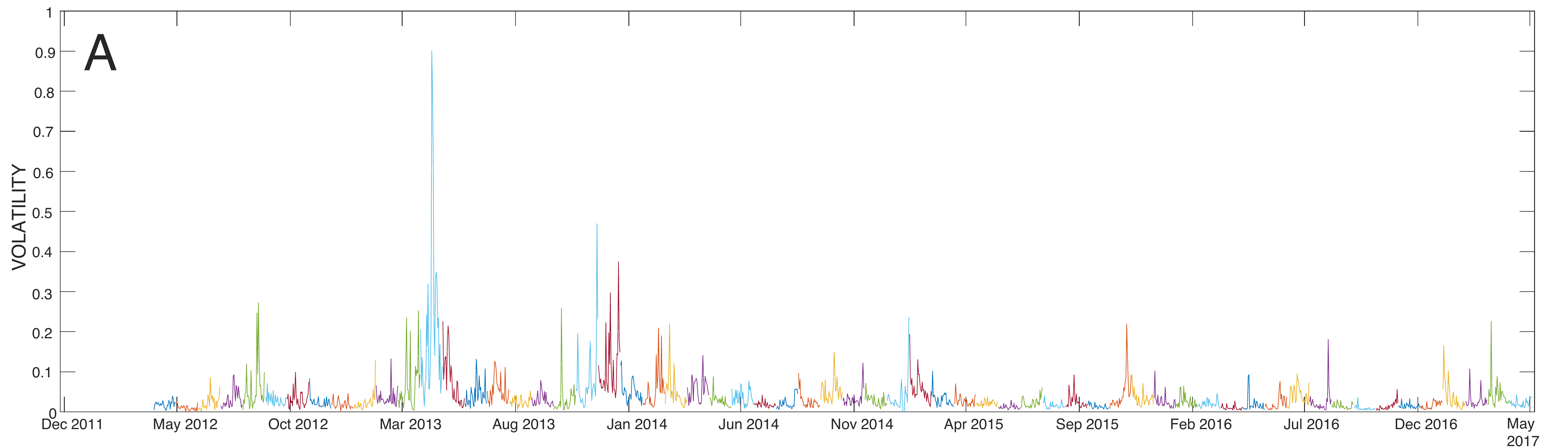}{}
\piclabC{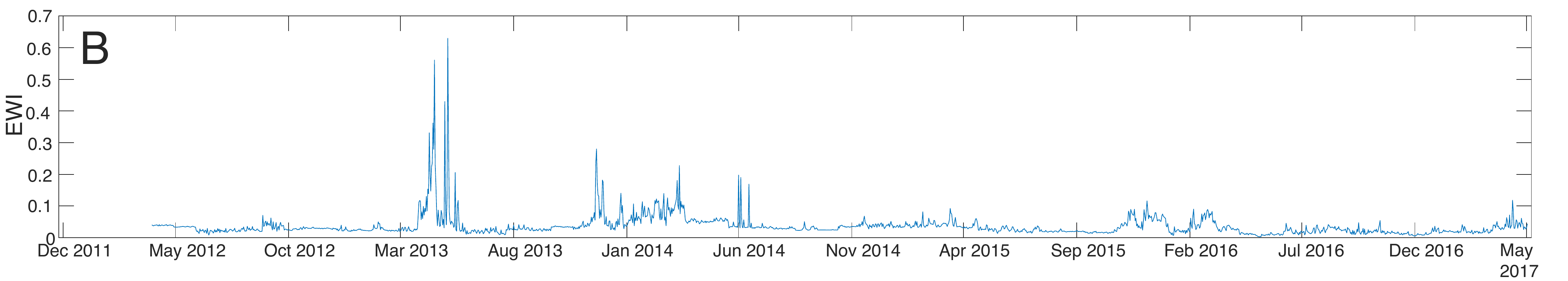}{}
\piclabG{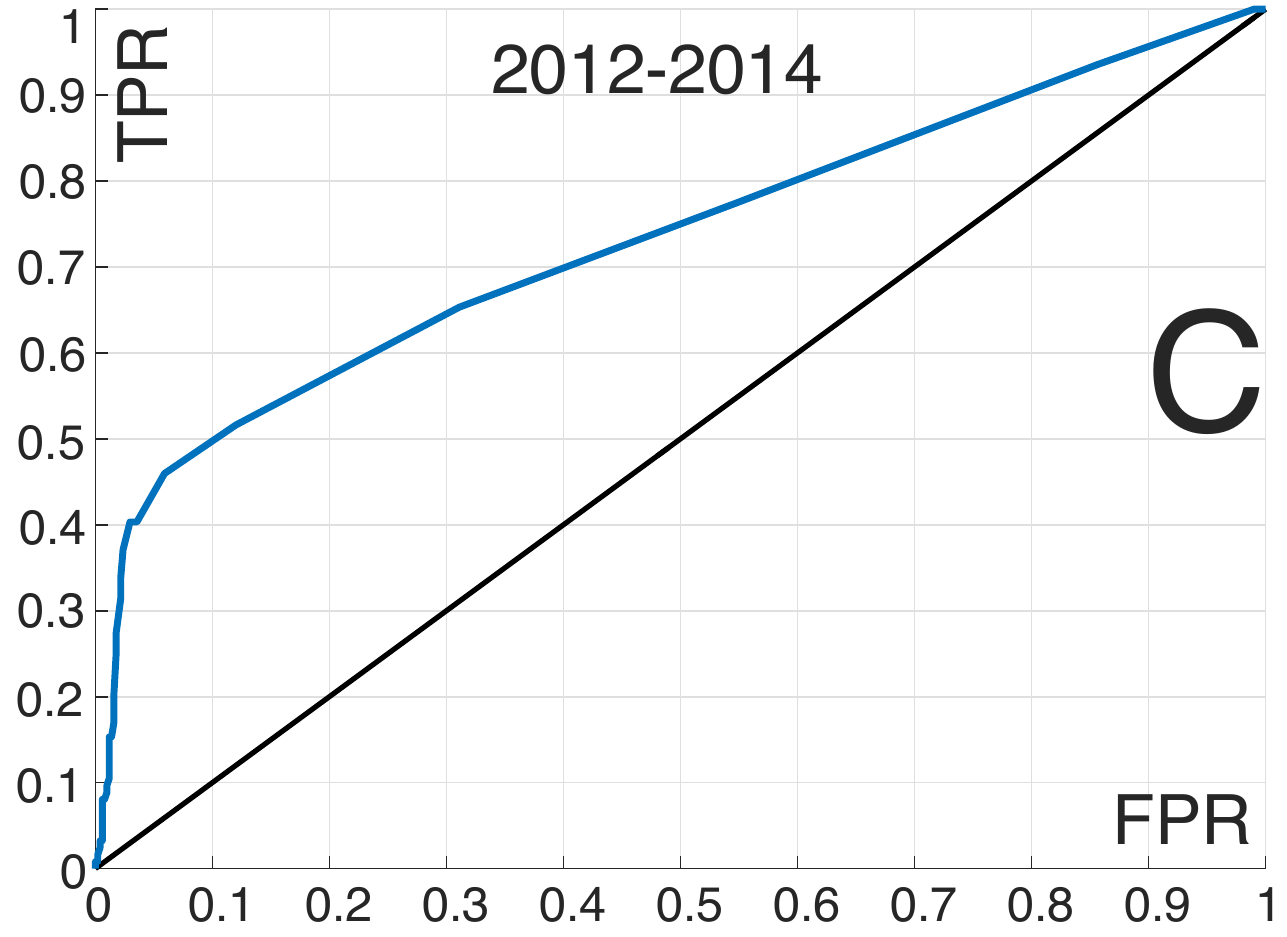}{}
\piclabG{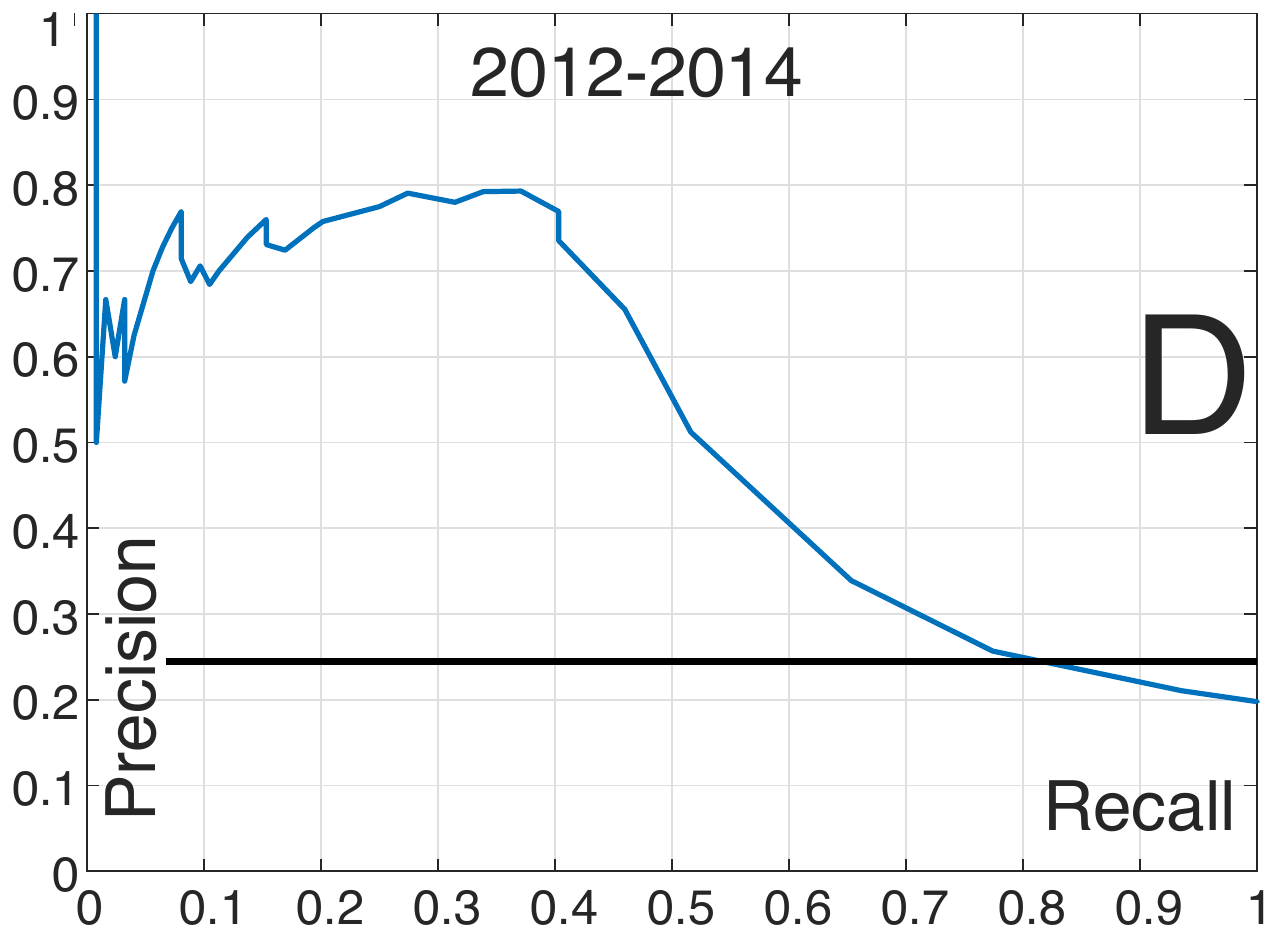}{}
\piclabG{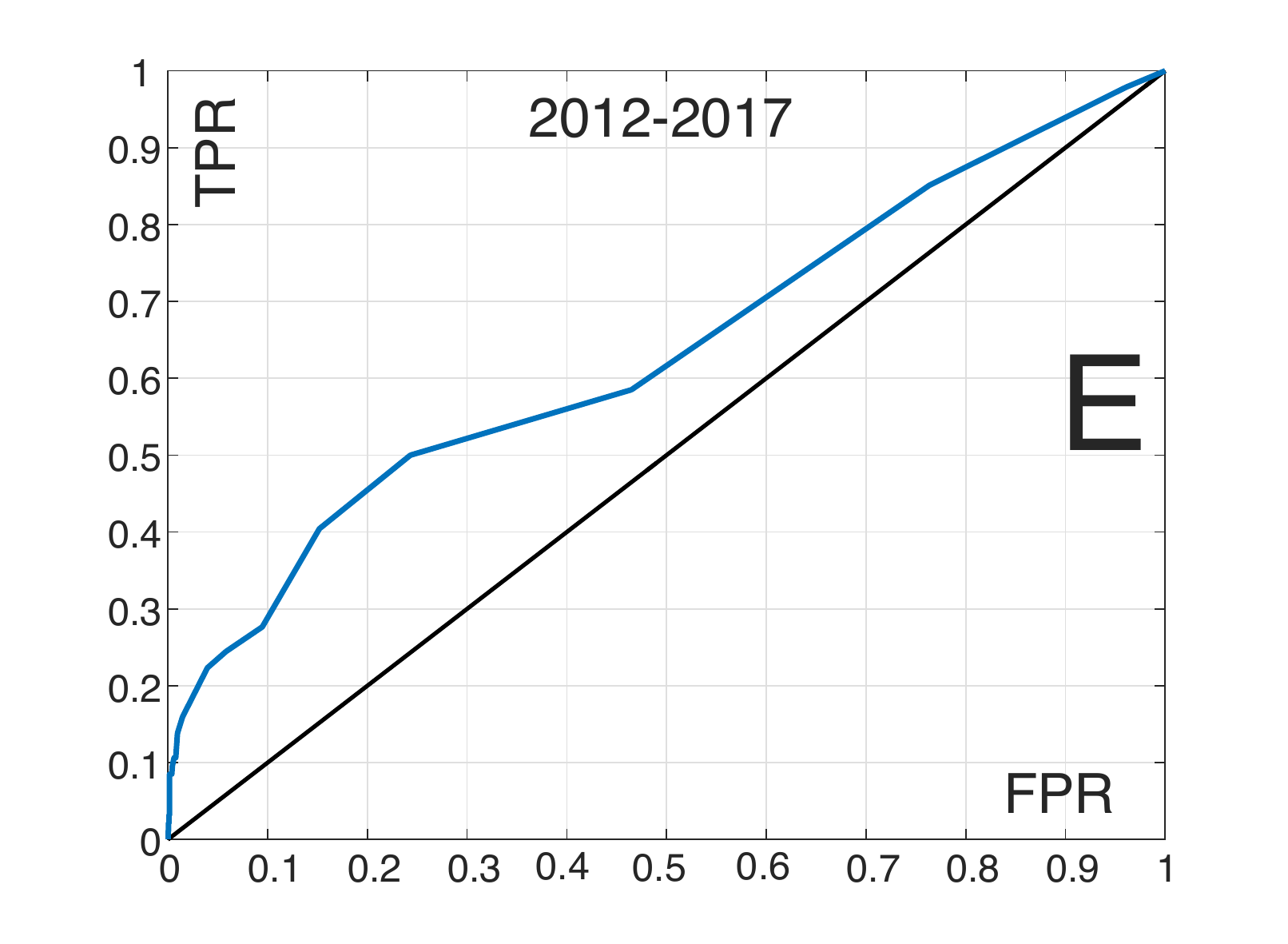}{}
\piclabG{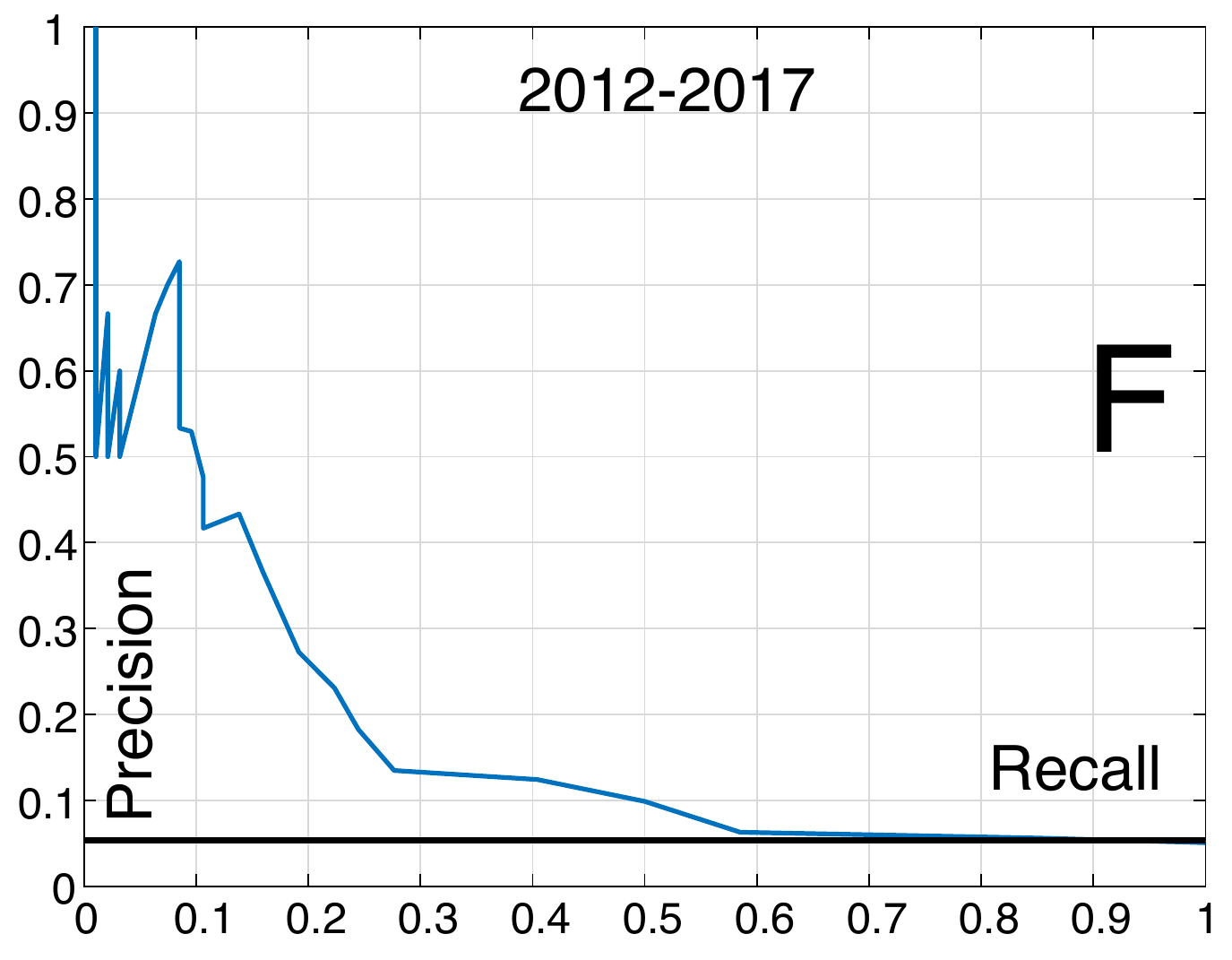}{}
\end{center}
\caption{
\textbf{Panel A}: Hold-out (GK daily volatility) segments color coded for each month. For each hold-out segment of one month ($\Delta=30$ days) the learning is done on previous $M=5*30$ days of data, while evaluation is done on the one month period. Re-training happens after every 1-month segment in a rolling window scheme. 
\textbf{Panel B}: Early warning indicator as Linear Non-negative Auto-Regressed NMF signal (NMF-NLR learning $M=5$ month, hold out $\Delta=1$ month, k = 10 factors, $\delta=5$). In plots B-F, we plot ROC and PR performance prediction curves for next day $h=1$ (in blue color) vs random noise as a baseline (in black color) for different time segments and $\alpha=0.1$.
\textbf{Panel C}: ROC with AUC 0.73 for period 2012-2014.
\textbf{Panel D}: PR curve with AUC 0.51 for period 2012-2014.
\textbf{Panel E}: ROC curve with AUC 0.65 for period 2012-2017. 
\textbf{Panel F}: PR curve with AUC 0.2 for period 2012-2017.
}
\label{fig:EWS_grounTruth_PR_ROC}
\end{figure}

\textbf{3.2. Evaluation}\\
We use a \textbf{receiver operating characteristic} (ROC) curve that gives a prediction ability of a binary classifier as its discrimination threshold $\Theta$ is varied. The ROC curve is created by plotting the true positive rate (TPR) against the false positive rate (FPR) at various threshold settings $\Theta$. True positive rate is a proportion of true extreme events that were correctly classified as such $\# [\eta(.) \geq \Theta] / \# [\beta(.)=1]$, while the false positive rate is proportion of our predicted extreme events that were falsely classified as such $\#[\eta(.) < \Theta] / \# [\beta(.)=1]$. 
We compare the area under the ROC curve (AUC ROC) against a baseline for the random signal, where AUC ROC (RND) equals 0.5. Due to the fact that the extreme events are much less common than non-extreme events, we also use the area under \textbf{precision-recall} curve \cite{Davis2006} (PR). Note that recall is equivalent to the true positive rate and precision is defined as the proportion of our predicted extreme events which are indeed extreme $\# [\beta(.)=1] / \# [\eta(.) \geq \Theta ]$. We compare the area under the PR curve (AUC PR) against a baseline for the random signal, where AUC PR (RND) is denoted as $\epsilon$. 
Here $\epsilon$ is the fraction of events that have the positive ground truth label $\beta(\cdot)=1$. 
In Figure \ref{fig:EWS_grounTruth_PR_ROC} we show the EWI (plot B), along with volatility (plot A), ROC and PR performance curves (plots C-F). We observe that the EWI ($k=10$, $\delta=5$) in period 2012-2014  can predict future extreme events ($\alpha=0.1$, $h=1$) with the following performance ($TPR \approx 0.7$, $FPR \approx 0.4$) on plot C and (precision $\approx 0.8$, recall$ \approx 0.2$) on plot D. 
In the next section we analyze the sensitivity of prediction in more details.

\begin{figure}[h]
\begin{center}
\piclabD{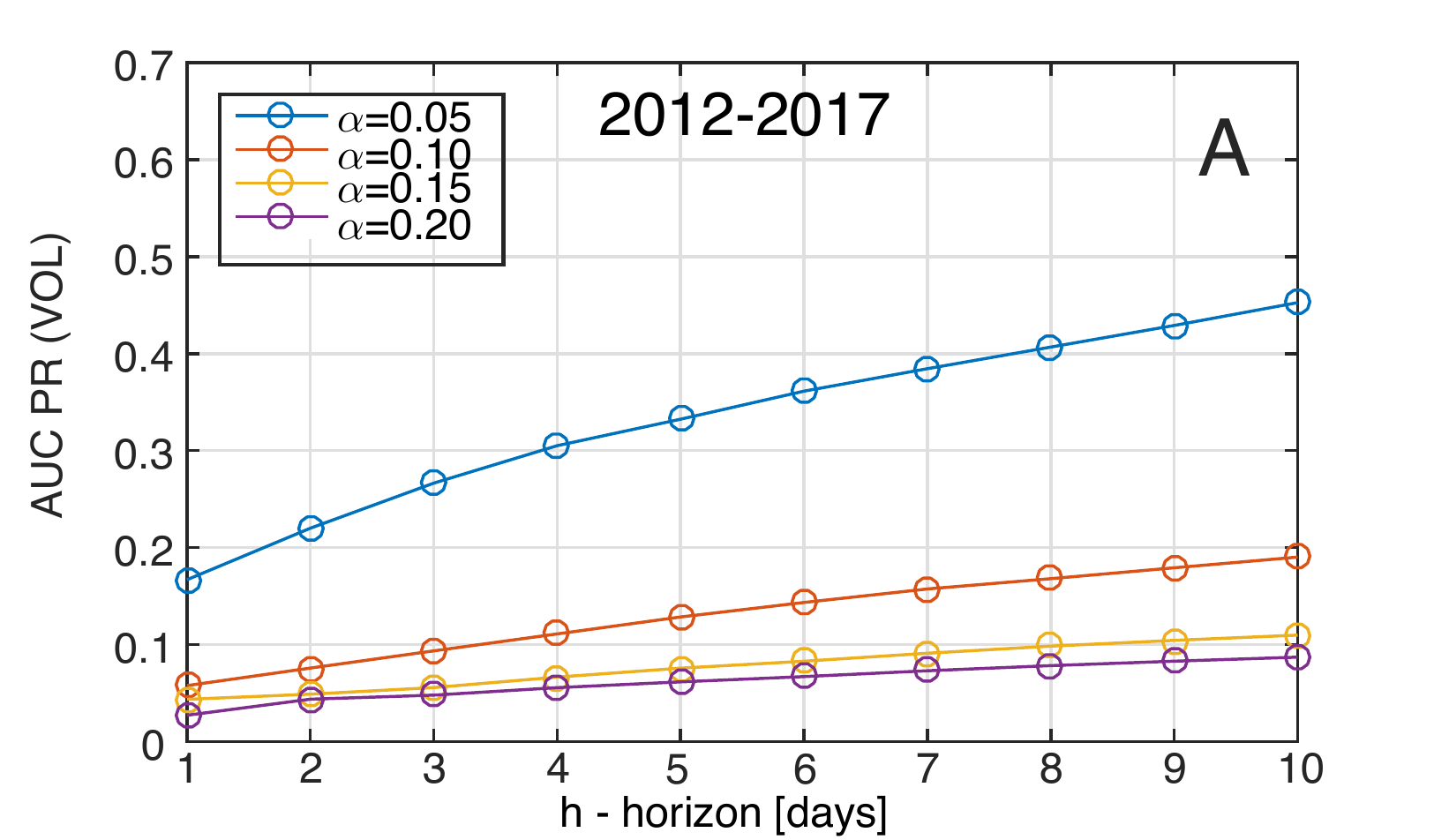}{}
\piclabD{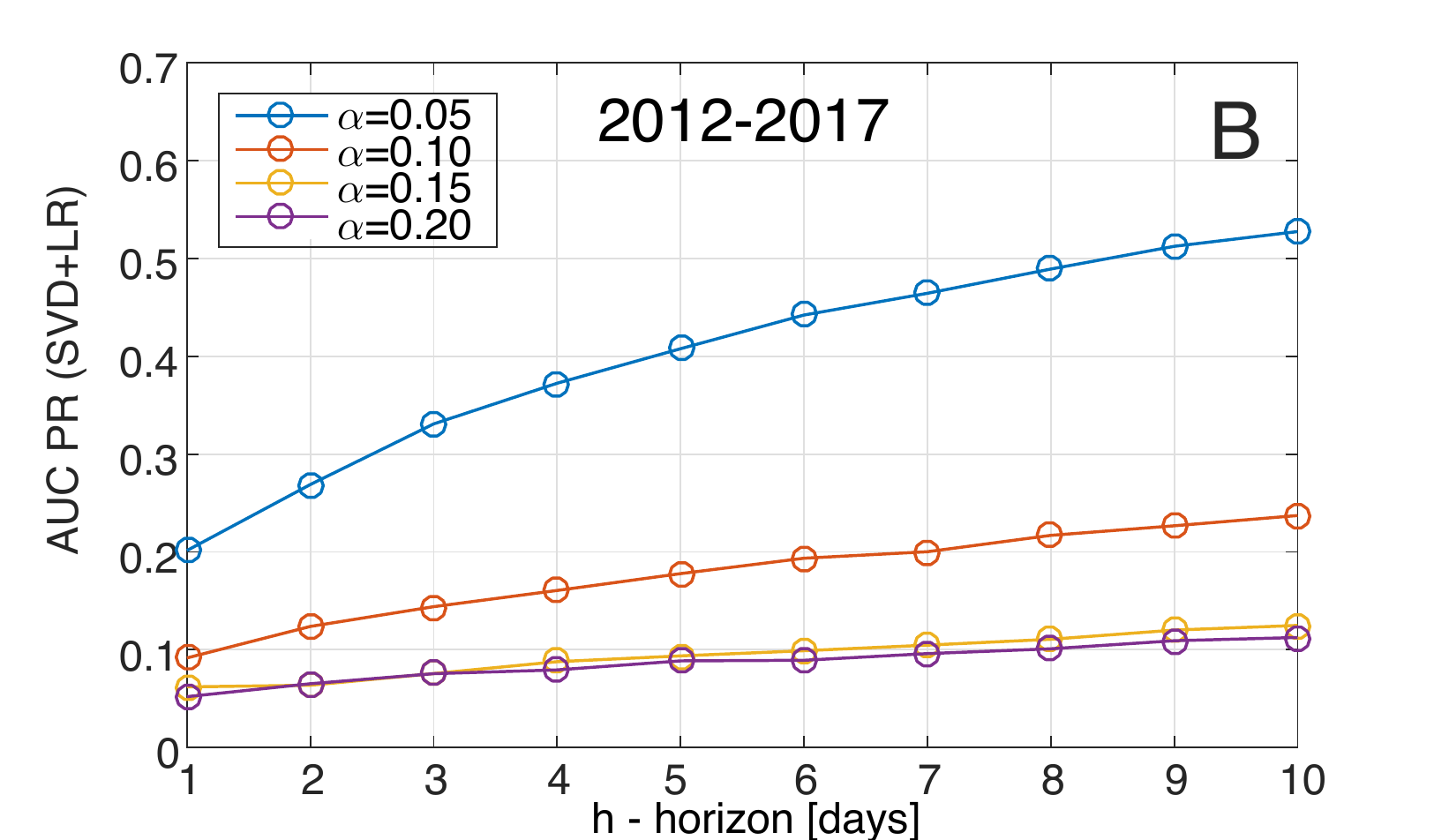}{}
\piclabD{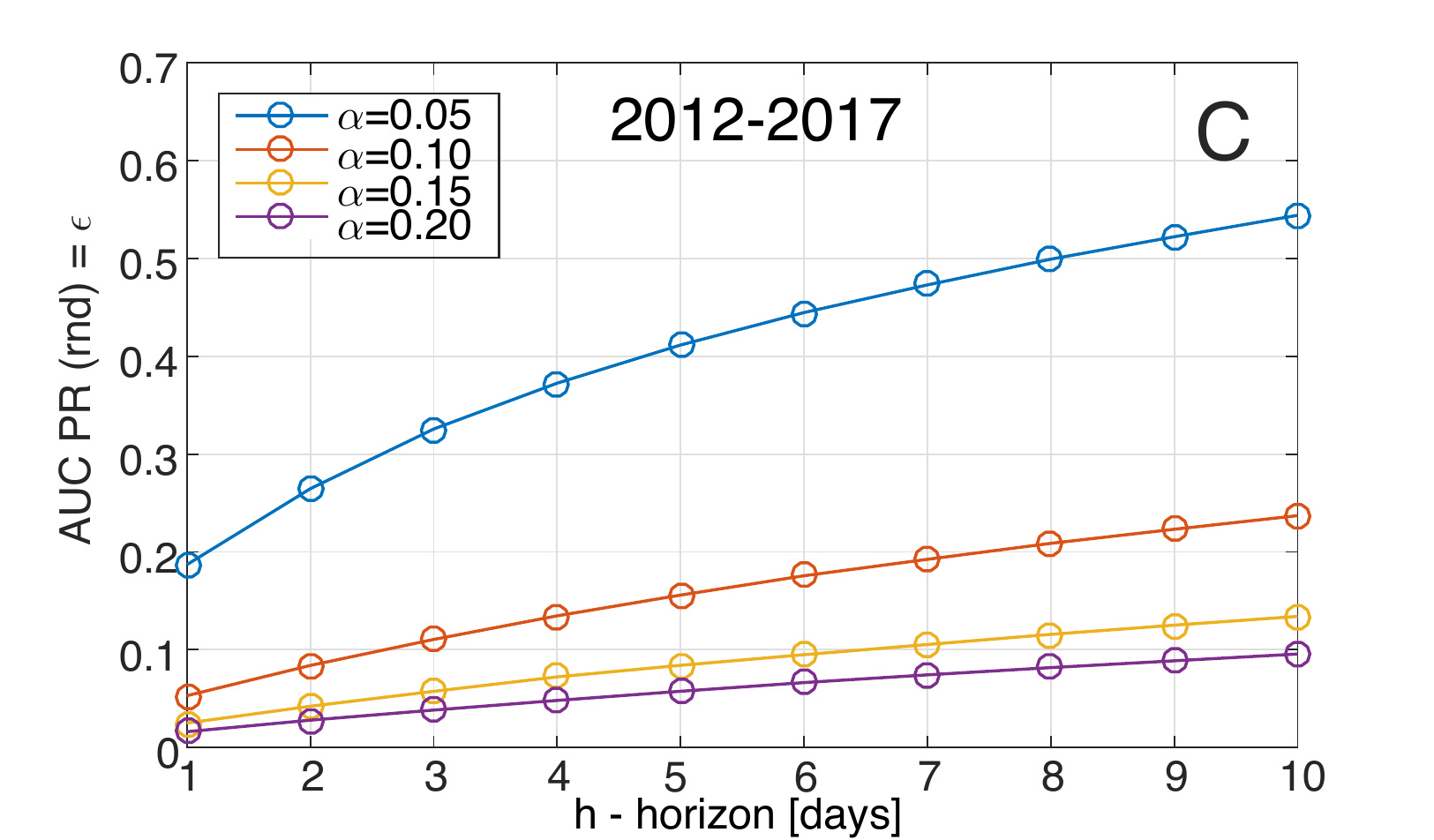}{}
\piclabD{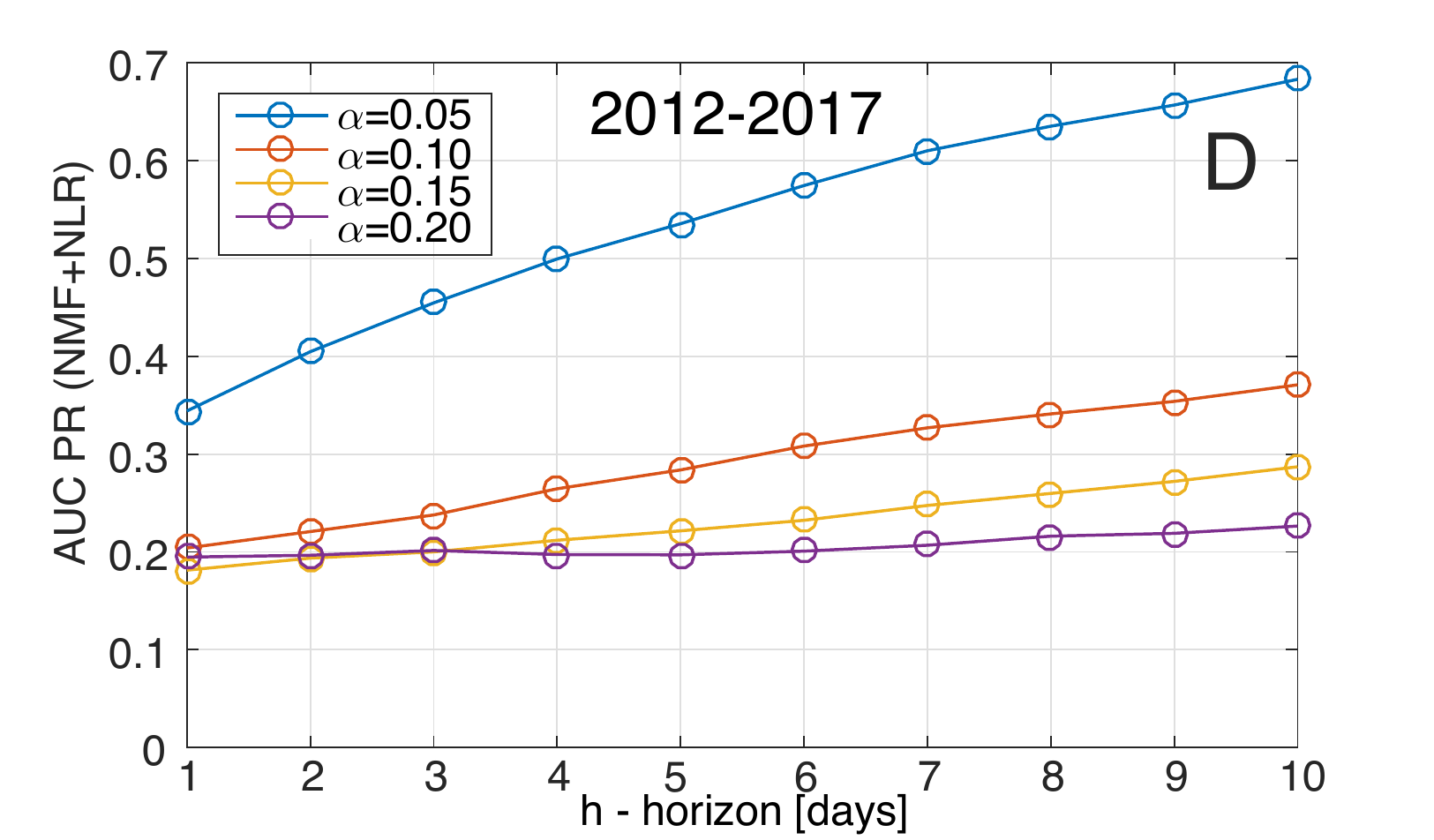}{}
\end{center}
\caption{ Area under PR (AUC PR) curve in period 2012-2017 for corresponding $\alpha$ (extreme event threshold) and $h$ (localization prediction horizon). \textbf{Panel A} AUC PR for BTC Blockchain volume. 
\textbf{Panel B}: AUC PR for time-series constructed with Ridge Linear Regression over low-dimensional representations, obtained with SVD.
\textbf{Panel C}: The ratio $\epsilon$ of extreme events for different $\alpha$ thresholds and horizons $h$. This ratio is equal to the AUC PR for random signal.
\textbf{Panel D}: AUC PR for early warning indicator, $k=10$ NMF factors and non-negative linear regression, $\delta=5$ regressive days.
}
\label{fig:PR_baselines}
\end{figure}

\textbf{3.3. Statistical and sensitivity analysis for EWI}\\
In the previous section we have showed the ROC and PR curves for fixed parameters: $k=10$ factors, $\delta=5$ regressive days and evaluation for predicting extreme event $\alpha=0.1$ within horizon of $h=1$ day. In this section, we make statistical and sensitivity analysis by providing the area under the curve statistics (AUC) over all possible $\Theta$ parameters and comparing it to the AUC of a random classifier. 
Note that prediction of extreme events within different localization horizons $h$ differs in prediction difficulty. E.g. prediction of extreme event happening at horizon $h=1$ days in future is more localized prediction than predicting the extreme event happening at next $h=5$ days in future. 
Furthermore, as we are dealing with different ratios of extreme events (imbalance dataset) only PR curves are used for sensitivity analysis for different horizons $h$. This is due to the fact that ROC curves are not sensitive as PR curves for skew imbalance ($\epsilon << 0.5$) in datasets \cite{Davis2006}. Sensitivity analysis for different levels of $\alpha$ thresholds, $\delta$ auto-regressive order parameters and $k$ number of NMF factors are all taken into consideration. 

In Figure \ref{fig:PR_baselines} panel A, we see the AUC PR performance for the the early warning indicator derived from a blockchain volume time series volume i.e. a total number of bitcoins in transaction networks at day $t$. 
In Figure \ref{fig:PR_baselines} panel B, we use low-dimensional features obtained from a singular value decomposition, along with linear regression as the second baseline for the EWI. This baseline is very similar to the Kondor et. al. study \cite{KondorBTC}.
In the case of both baselines, we observe that the AUC PR performance of the EWI increases as the localization length $h$ of the extreme event increases. This is in correspondence with our assumption that the predictability changes for different values of $h$. 

In the Figure \ref{fig:PR_baselines} panel C, we can see the performance of the random baseline, which increases with the localization length $h$. As the prediction segments become larger so does the probability of the occurrence of the extreme event by chance. In Table \ref{tab:auc_performance} we show the numerical values for different baselines and different values of parameter $\alpha$. We observe that the proposed inferred signal (NMF+NLR) has the highest prediction performance. 
In Table \ref{tab:auc_performance}, part A, we also show the difference between the AUC performance of EWI and AUC performance of a random baseline (RND).
We observe that on average it is easier to predict less extreme event - smaller values of parameter $\alpha$. 
In a case when the prediction is only based on the features from a current day ($\delta=1$, see Table \ref{tab:auc_performance}, part B) the predictions of extreme events ($\alpha=0.1$, $\alpha=0.15$ and $\alpha=0.2$) significantly drops with respect to $\delta=10$ case. This shows that the historical auto-regressed terms ($\delta>1$) are important for predictions.
In general, sensitivity analysis shows that results are also relatively stable for different parameters of $k$ (Table \ref{tab:auc_performance}, part C). 
However, more in-depth analysis of the embedding dimensionality $k$ was out of the scope of the current work and is left for future work. 

\begin{table}[h]
\begin{center}
\begin{tabular}{l c c c c c}
& Indicator ($h=1$)         & $\alpha=0.05$ & $\alpha=0.10$ & $\alpha=0.15$ & $\alpha=0.20$    \\
\hline
\hline
& AUC PR (VOL) & 0.167 & 0.057 & 0.043 & 0.027 \\
A & AUC PR (RND) & 0.186 & 0.053 & 0.025 & 0.016 \\
 & AUC PR (SVD+LR)[$k$=10,$\delta=5$] & 0.201 & 0.092 & 0.062 & 0.052 \\
& AUC PR (NMF+NLR)[$k$=10,$\delta=5$] & 0.344 & 0.204 & 0.181 & 0.195 \\
\hline
\hline
B & AUC PR (NMF+NLR)[$k$=10,$\delta=1$] - AUC PR (RND) & 0.172 & 0.064 & 0.033 & 0.021 \\
& AUC PR (NMF+NLR)[$k$=10,$\delta=10$] - AUC PR (RND) & 0.164 & 0.129 & 0.130 & 0.127 \\
\hline
\hline
C & AUC PR (NMF+NLR)[$k$=5,$\delta=5$] - AUC PR (RND) & 0.149 & 0.137 & 0.176 & 0.171 \\
& AUC PR (NMF+NLR)[$k$=20,$\delta=5$] - AUC PR (RND) & 0.194 & 0.143 & 0.165 & 0.150 \\
\end{tabular}

\caption{Performance of different early warning indicators for segment 2012-2017. \textbf{Part A:} AUC PR performance for different indicators: (i) VOL - volume in transaction graphs, (ii) SVD+LR - Linear regression over low-dimensional representations of Singular Value Decomposition, (iii) NMF+NLR - Non-negative linear regression over low-dimensional representations of Non-negative Matrix Factorization, (iv) RND - random classifier.
\textbf{Part B:} Performance for different time order legs $\delta$. \textbf{Part C:} Performance for dimensionality of representations $k$
}
\label{tab:auc_performance}
\end{center}
\end{table}





\begin{figure}[h]
\begin{center}
\piclabH{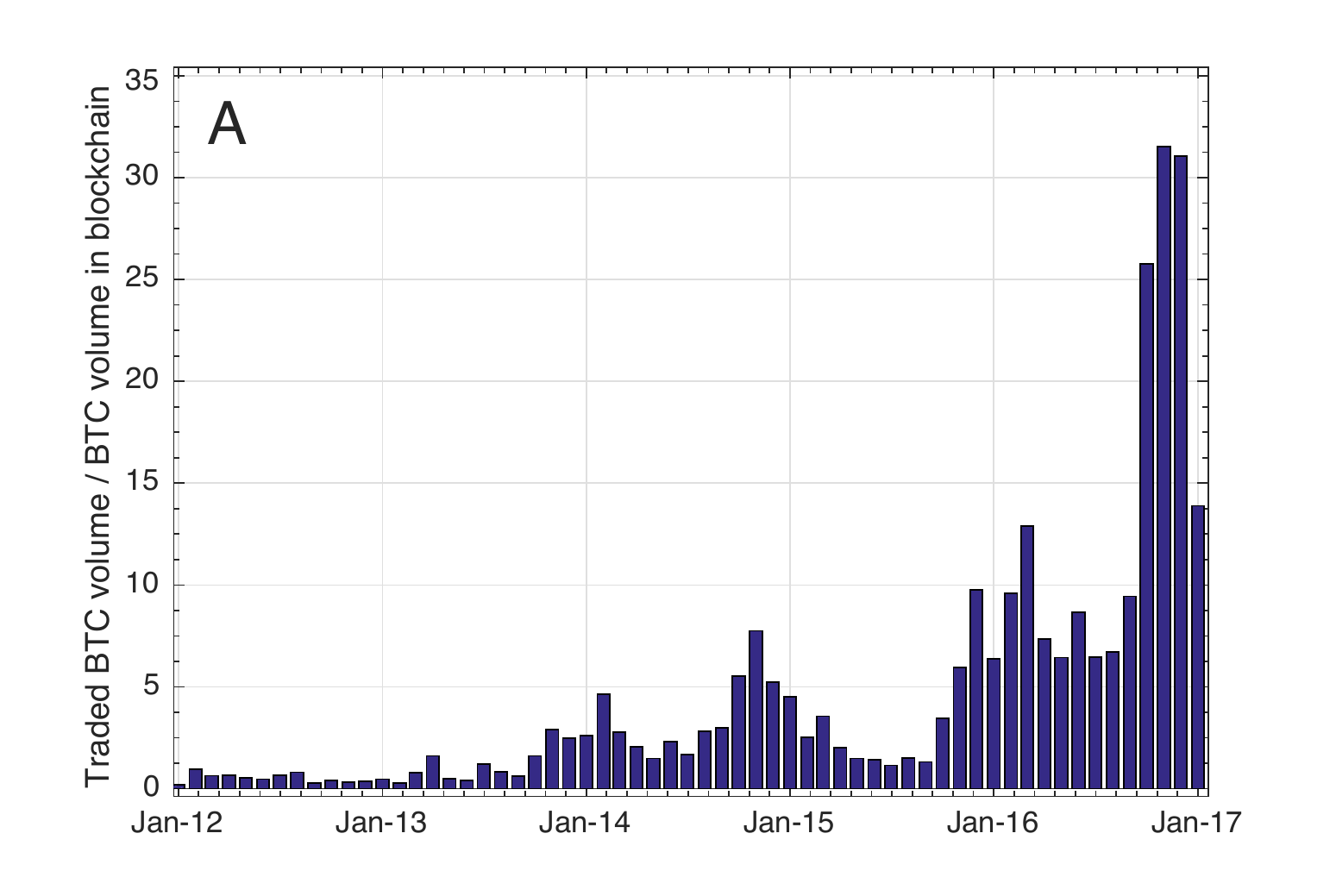}{}
 \piclabH{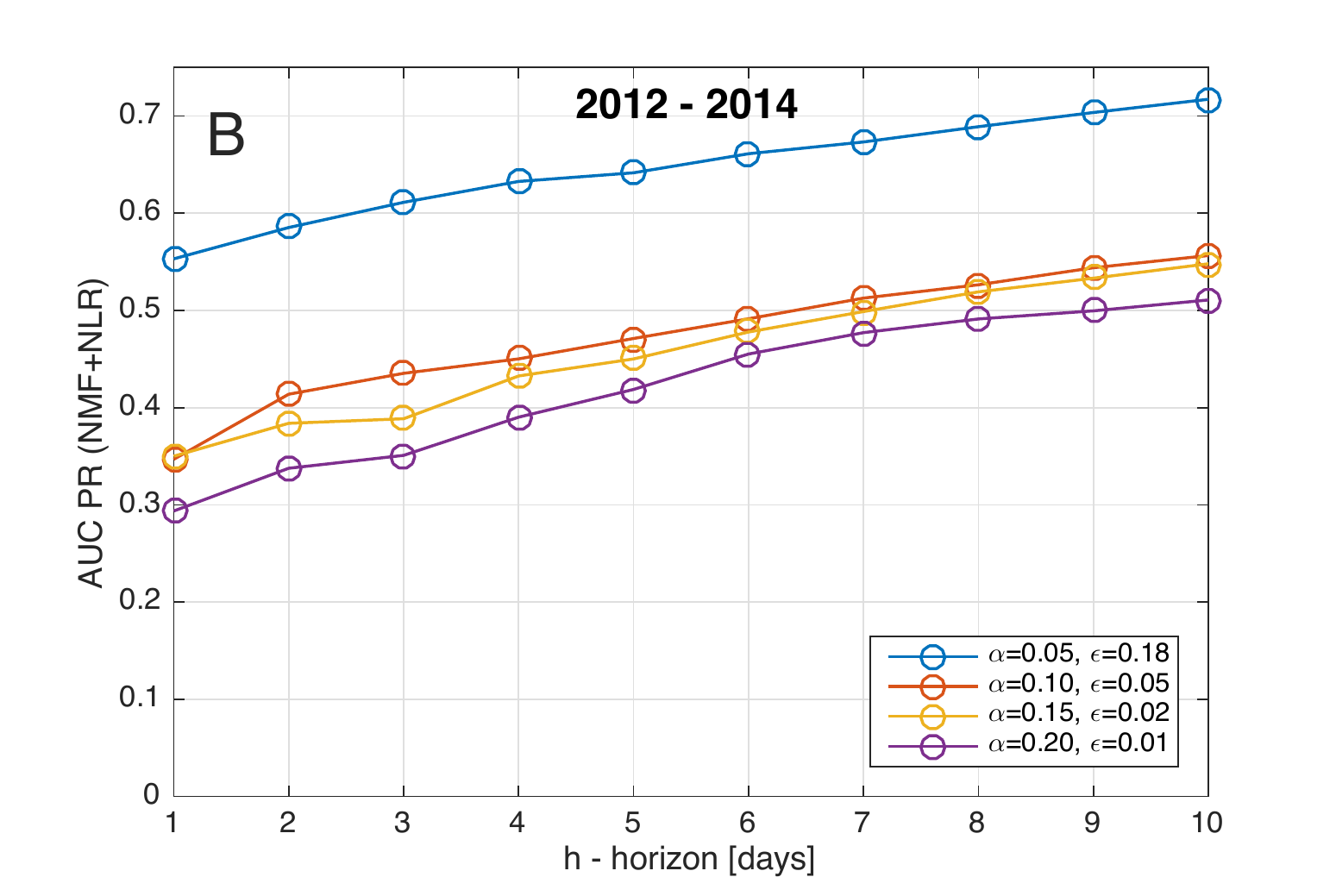}{}
\end{center}
\caption{ \textbf{Panel A:} The ratio $V_M(t)/V_B(t)$ of total traded bitcoin volume on major exchange markets (bitfinex, bitflyer, bistamp, btcchina, coinbase, lakebtc, mtgox, okcoin and others) and total volume in our Blockchain transaction graphs. In November and December 2017 the total traded volume in units of bitcoins exceeded 160 million.  
\textbf{Panel B:} Area under PR curve for early warning indicator with $k=10$ NMF factors and non-negative linear regression, $\delta=5$ regressive days and different $\alpha$ conditions for extreme events at different localization horizons $h$ in future.
for period 2012-2014.
}
\label{fig:market_blockchain}
\end{figure}

\section{Discussion and conclusion}
In this paper we analyze the performance of early warning indicators for extreme future volatility in two different time periods: (i) 2012-2014, and (ii) 2012-2017. We observe that the performance during the first (shorter) period up to 2014 is better compared to the performance over the entire period analyzed. On one hand, the ROC AUC and the PR AUC are 0.73 and 0.51 respectively for the period between 2012-2014, while, on the other, for the entire period (2012-2017), the ROC AUC and the PR AUC are 0.65 and 0.2 respectively (See Fig. 2. C-F). To better understand the differences in model performance between these two periods, we study the changes in the ratio of (i) total market exchange volume in Bitcoin $V_m(t)$ and (ii) the Bitcoin volume in the transaction graphs that we analyze $V_b(t)$. $V_m(t)$ includes all Bitcoin exchange transactions on the following exchanges: Bitfinex, Bitflyer, Bistamp, BtcChina, Coinbase, LakeBtc, MtGx, OkCoin, and others). We find that the ratio $V_m(t) / V_b(t)$ increases tenfold after 2014, from a maximum value of 3 during the period 2012-2014 to over 30 in 2017. This implies that there is a significant overwhelming interest in Bitcoin as a speculative investment asset, compared to its use as payment mechanism for purchasing and selling goods and services, represented by the number of transactions on the transaction graphs that we have analyzed. Hence, due to this dynamics, there is a significant deficiency in information obtained from the transaction graphs relative to the information contained in speculative trading  or using Bitcoin as short-term investment asset. This trend is due to the slow maturing of Bitcoin as a payment method and the skepticism of its wide adoption due to lack of regulation and fear of significant loss in value due to electronic theft of Bitcoins or extreme volatility. Our hypothesis is that the transaction graphs or the relational aspect of Bitcoin will inform more about future volatility and can become an important early warning signal for ensuing volatility once Bitcoin becomes more mature payment method in trades of gods and services, which is an interesting topic for future research. \\



\textbf{Acknowledgement and contribution}\\
\small{Thanks to students Gr{\"u}ner Maximilian, Weingart Nino, Riesenkampf Heiki 
for help in processing blockchain data.
The work of N.A.F. has been funded by the EU Horizon 2020 SoBigData project under grant agreement No. 654024. 
All authors contributed to the writing and editing of the manuscript.
N.A.F. performed computational modeling and experiments. 
D.T. performed computational modeling and design of research. 
M.P. and Z.C. were involved in data processing and analysis.
I.V. was involved in financial analysis and interpretation of results.}


\section*{Appendix}
In order to solve the following non-convex optimization problem $
\underset{\textbf{H,W} \geq 0}{\textrm{min}} ||\textbf{X}-\textbf{WH}||_{2,1} + \lambda ||\textbf{H}||_{2,1}$
where $||.||_{2,1}$ denotes the $L_{2,1}$ matrix norm.
First we randomly initialize the matrices $\textbf{H,W}$
then iteratively fix one of the matrices (W,H) and perform the update step on another matrix. The procedure is repeated until the convergence. 
We use the following updates\cite{RNMF}:
$\textbf{H}_{k,i} = \textbf{H}_{k,i} \frac{(\textbf{W}^T \textbf{X} \textbf{D}_1)_{k,i}}{ (\textbf{W}^T \textbf{W} \textbf{H} \textbf{D}_1 + \lambda \textbf{H D}_2)_{k,i}}$,
$
\textbf{W}_{j,k} = \textbf{W}_{j,k} \frac{(\textbf{X D}_1 \textbf{H}^T)_{j,k}}{ ( \textbf{W H D}_1 \textbf{H}^T )_{j,k}}$,
where $\textbf{D}_1, \textbf{D}_2$ are diagonal matrices defined as:
$(\textbf{D}_{i,i})_1 =  1 / \sqrt{\sum_j (\textbf{X}-\textbf{WH})^2_{j,i}}$ , $(\textbf{D}_{i,i})_2 =  1 / \sqrt{\sum_j \textbf{H}^2_{j,i}}.$

\end{document}